**PERSPECTIVE**

# Hyperspectral imaging solutions for brain tissue metabolic and haemodynamic monitoring: an updated perspective

**Luca Giannoni[1]*, Frédéric Lange[1] and Ilias Tachtsidis[1]**

[1] Department of Medical Physics and Biomedical Engineering, University College London, London WC1E 6BT, United Kingdom

*Author to whom any correspondence should be addressed.

**E-mail:** l.giannoni@ucl.ac.uk

## Abstract

Since the publication of our review article "*Hyperspectral imaging solutions for brain tissue metabolic and hemodynamic monitoring: past, current and future developments*" in 2018, the technological and applicational landscape of the use of hyperspectral imaging (HSI) in brain sciences has evolved and transformed significantly. The number of studies and works where HSI has been deployed in its many forms to map and monitor the haemodynamic and metabolic states of cerebral tissues have grown exponentially, to such a point where an update on the current state of the art is timely, and we believe would be desirable for both long-term experts in the field, as well as for any new researcher approaching it for the first time. In this commentary, we provide a renewed perspective on the newest and latest developments in brain haemodynamic and metabolic monitoring with HSI over the past eight years. Our hope is that even greater breakthroughs and broader, more numerous novel applications will come forward in the future for the technology, that may benefit from this new overview, as they did from the original one.

## 1. Introduction

In March 2018, we first published, in a special issue on hyperspectral imaging (HSI) of the *Journal of Optics,* a review article titled "*Hyperspectral imaging solutions for brain tissue metabolic and hemodynamic monitoring: past, current and future developments*" [1]. Back then, HSI was a promising, label-free, optical imaging modality that was just starting to venture effectively into life sciences and biomedical applications (the very first, comprehensive review articles on medical HSI are dated to early 2010s [2,3]), originating distantly from more established and previously explored fields [4] such as remote sensing [5], agriculture [6], geology [7], and food processing [8] and quality control [9]. In particular, its deployment as a monitoring and mapping tool for brain sciences was at its inception, with just a limited number of studies available in the literature (e.g., in our 2018's review article, we focused primarily on merely 10-11 spotlighted publications), where quantitative (or semi-quantitative) tracking of selected biomarkers for intrinsic oxygenation and metabolic contrasts had been successfully presented. The vast majority of these works were at preclinical level, on *in vivo* animal models or *ex vivo* samples, with only a few including *in-vivo* clinical applications on humans. Similarly, almost in the entirety of the reported cases, in-house and custom-made systems were employed, with very little references on the use of commercial HSI cameras or similar off-the-shelf devices. The latter fact we attributed to the premature state of the market and industry for hyperspectral technologies, where biomedical and biological applications were still not the primary target. Thus, performances and readiness of commercial HSI solutions for haemodynamic and metabolic monitoring of the brain was lagging behind at the time with state-of-the-art, research laboratory instrumentation [10].

As of 2026, the landscape has now significantly and promisingly evolved [11,12]. New technologies, novel approaches and far more numerous applications, bridging the preclinical with the clinical realms, have blossomed in the field of haemodynamic and metabolic mapping of the brain since the publication of our review article, where we discussed past, current and future trends. Most of those future trends have now started to concretely materialise in a larger number of works, thus we believe it is the right time for a due, updated perspective on the state-of-the-art and onto any new, forthcoming directions that the field is seemingly taking.

As we did for our original article in 2018, our prospect is also for this new and updated commentary to be of use and inspiration for any experienced -as well as new- researcher and scientist who was in the past, or currently is, or will be looking forward to contributing to such a novel and flourishing area of investigation.

## 2. Recent updates on HSI of the haemodynamic and metabolic states of the brain

In our original, 2018 review article [1], the majority of the works we presented on the use of HSI to monitor the haemodynamic and metabolic states of the brain had their methodological foundations lying on the precursory achievements and long-established approaches from the fields of functional near-infrared spectroscopy (fNIRS) [13,14] and diffuse optical tomography (DOT) [15,16]. Previous HSI attempts at measuring the intrinsic optical signatures of known biomarkers of brain oxygenation, haemodynamics and metabolism in the brain, such as oxy- ($HbO_2$) and deoxyhaemoglobin (HHb) [17], or cytochrome-co-oxidase (CCO) [18], were essentially based on extending the 1D or topographic imaging methods of fNIRS and DOT to the 2D, widefield imaging domain of HSI [19,20].

In particular, the modified Beer–Lambert law (MBLL) [21] was primarily used by almost every single publication we reviewed in 2018, to calculate relative changes in the concentrations of $HbO_2$ and HHb, as well as oxidised (oxCCO) or reduced CCO (redCCO), from measurement of light attenuation at multiple wavelength bands in the visible and NIR range (for each pixel of the hyperspectral images). Based on the latter, we identified three major categories of methods for the quantitative mapping of haemodynamic and metabolism in the brain: (I) measuring spatially localised variations in the concentration of $HbO_2$ and HHb as direct biomarkers of oxygenation and haemodynamics, yet indirect indicators of brain metabolism; (II) the quantification of brain oxygen metabolism, via quantitative mapping of the cerebral metabolic rate of oxygen ($CMRO_2$) [22,23]; and (III) the quantitative mapping and targeting of direct biomarkers of brain tissue energetics (i.e., oxidative metabolism), mainly the mitochondrial redox states of CCO, as well as the autofluorescence emitted by cytoplasmatic metabolites, such as reduced nicotinamide adenine dinucleotide (NADH) and oxidised flavin adenine dinucleotide (FAD) [24].

### 2.1. The future is now: new HSI approaches

As much as the MBLL-based, original approaches still lie at the core of most of the newest (to date) applications of HSI to cerebral haemodynamics and metabolic monitoring since our 2018 review [25–30], novel trends have now focused on how to better implement such methodologies and extend their applications within brain sciences.

In particular, latest studies have now concretely fulfilled what we envisioned in 2018 as a future perspective -i.e., the HSI of cerebral cellular energetics as a novel way to directly map and quantify brain metabolism- and transformed it into actual reality. For instance, HSI targeting of the redox states of CCO in the brain have now become an established metabolic monitoring approach, whilst at the time of our previous review this was still at a transitional state from its 1D/topographic counterpart, i.e., broadband NIRS (bNIRS) [18,31]. First, investigated with the use of Monte Carlo (MC) digital modelling [32,33], and then validated on realistic, dynamic optical phantoms [34,35], the technique is now consistently applied *in vivo*: Giannoni *et al.* [34] demonstrated the mapping and simultaneous monitoring of $HbO_2$, HHb and oxCCO on the exposed cortex of mice during various states of hyperoxia, hypoxia and anoxia with an in-house, spectral scanning system (Fig. 1a), whereas Caredda *et al.* [36,37] successfully used a commercial snapshot camera to translate the achievement in the clinics, mapping functional changes in $HbO_2$, HHb and oxCCO intraoperatively on humans during neurosurgery (Fig. 1b).

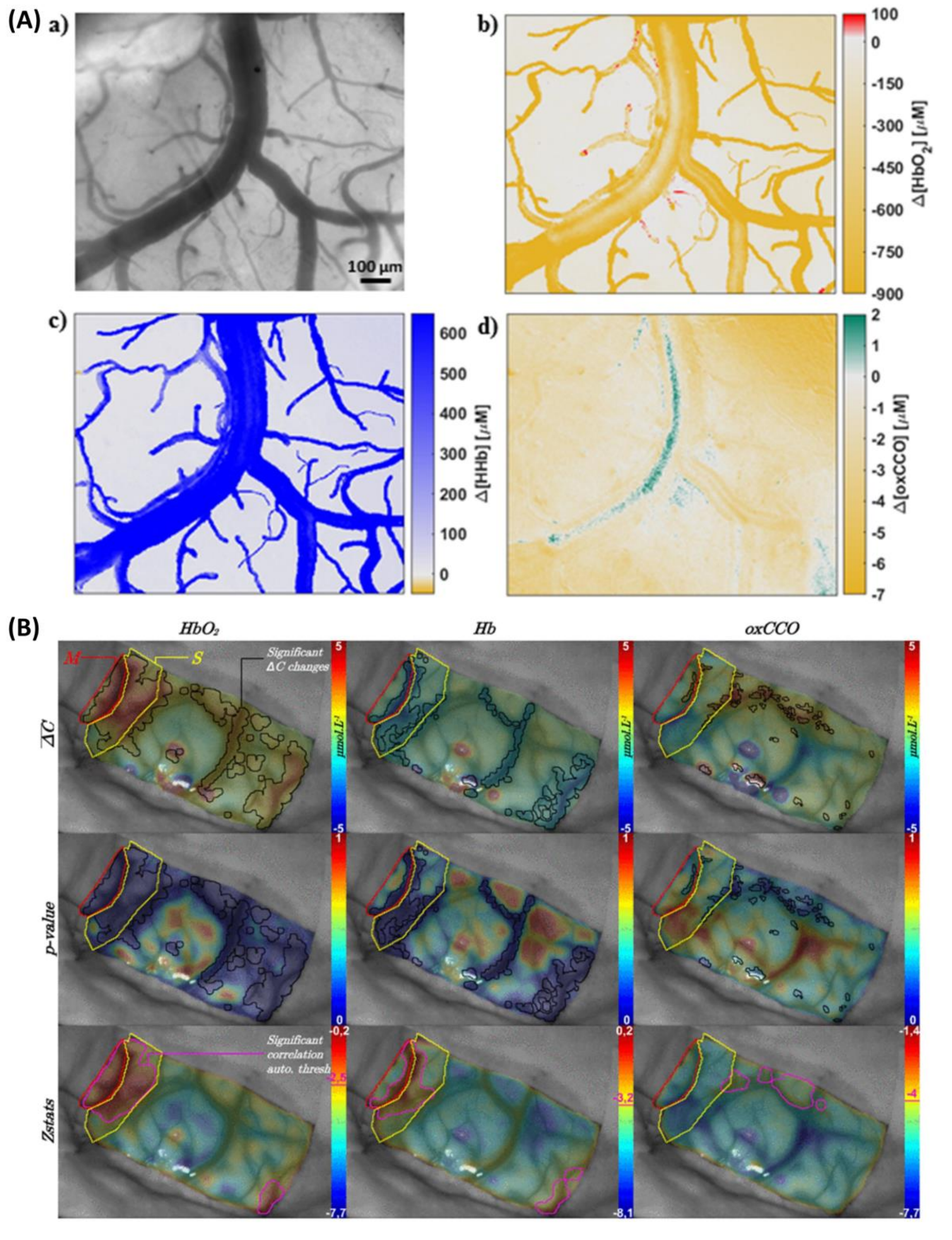

**Figure 1.** (A) Example of preclinical hyperspectral data of real-time monitoring and quantification of the haemodynamic (via haemoglobin oxygenation) and metabolic responses (via oxCCO) of the exposed brain cortex of a mouse to cerebral anoxia. All the three functional maps (b, c, d) show microvascular-level resolution (down to about 20 µm) over a field-of-view (a) of ~0.9x1.1 mm$^2$ [34]. (B) Example of intraoperative functional HSI brain mapping on awake patients during neurosurgery, demonstrating quantification of the haemodynamic and CCO contrasts for intraoperative identification of eloquent brain areas, using a commercial hyperspectral camera. The coloured bar represents the scale of variations of haemoglobin and oxCCO concentration changes in µMol·L$^{-1}$. Letters M and S indicate motor and sensory areas, respectively [36]. Reproduced with permission.

Similarly, HSI of fluorescent biomarkers of metabolism in the brain also started to advance from the solely *in vitro* and *ex vivo*, proof-of-concept studies we reviewed in our 2018 article into a number of applications *in vivo* [38,39], both preclinically on animal models [40] and clinically during surgery [41,42]. Although implementation of HSI targeting endogenous autofluorescence (such as generated by NADH and FAD) has mostly remained limited to cells and tissue samples (despite recent attempts to renewed translation [43]), potentially due to limitations in signal-to-noise ratio (SNR) and detectability *in vivo*, novel endogenous biomarkers have nonetheless emerged with significant potentials in clinical neuronavigation. Among these, the predominant approach is the HSI detection and mapping of 5-aminolevulinic acid (ALA)-induced fluorescence from protoporphyrin IX (PpIX) [44]. PpIX is a naturally-occurring, fluorescent precursor to haemoglobin and cytochromes, that can be excited at around 405 nm to emit dual peaks of fluorescence in the red, at 620 and 634 nm [45], with the ratio between peaks being primarily dependent on concentration and microenvironment [46,47]. Ingestion of 5-ALA as exogenous agent leads to temporary enhancement of PpIX concentration in metabolically active tissue, which makes it particularly suited to target tumoral cells, such as gliomas in the brain [47]. Thus, the use of HSI to map and quantify 5-ALA-induced PpIX fluorescence has now been actively implemented as a novel intraoperative approach for tumour identification and boundary delimitation in neurooncological surgery [48–51]. D'Alessandro *et al.* [52] recently demonstrated noteworthy accuracy in the quantification of PpIX concentrations in *ex vivo* glioma biopsies with HSI, when compared to reversed-phase liquid chromatography coupled to mass spectrometry (LC–MS), as a benchmark method to deliver precise concentration estimates.

### 2.2. From the laboratories to the markets: new HSI technologies

Significant progresses in the use of HSI to monitor cerebral haemodynamics and metabolism have also been made possible thanks to substantial advancements and recent developments in new hyperspectral technologies. At the time of our 2018 review article, spectral scanning methods were the most commonly adopted in preclinical *in vivo* studies, the majority of which were based on the use of conventional white light sources (such as lamps) coupled with motorised filter wheels for wavelength selection. Nowadays, more sophisticated illumination and filtering methods have been largely adopted, both preclinically and clinically: among these, we can mention the use of supercontinuum lasers [34,53,54], often coupled with acousto-optical tuneable filters (AOTFs) [55,56], or the use of high-power, incoherent sources, such as laser-driven plasma sources (LPS) [43,57], as well as the implementation of liquid crystal tuneable filters (LCTFs) [58–60]. All these, more advanced instruments allowed users to increase the speed of hyperspectral acquisition (down to the scale of tens of seconds), enhance the quality of the SNR in the images (thanks to higher outputs in the illumination), obtain finer spectral resolution (e.g., below 5-10 nm, full-width half maximum; FWHM), and to extend the number of selectable spectral bands (above ten and up to hundreds of bands).

In parallel to in-house, laboratory equipment, commercial HSI instrumentation has begun to venture towards biomedical applications with recent improvements in sensor performances (spatial resolution, format, number of wavelength bands and dynamic range, above all), with particular advantages in clinical environments, thanks to their fast acquisition times, compactness and ease-of-use. Both state-of-the-art commercial linear scanning and snapshot cameras have found large applicability in interventional and neuronavigational settings, primarily in intraoperative cerebral functional monitoring [36,37] and in glioma resection [61–64]. Furthermore, industrial endeavours are now moving towards fully translating biomedical HSI systems for brain applications onto the markets: notable examples are Hypervision Surgical [65], Imec [66], Headwall [67] and Diaspective Vision [68].

Finally, promising trends towards multimodal combination of HSI with other optical imaging techniques are currently being investigated to provide more comprehensive overviews into cerebral physiology as a whole. In our 2018 review, we already highlighted works showcasing the combination of HSI with laser speckle contrast imaging (LSCI) to add complementary, spatial information on cerebral blood flow to $HbO_2$ and HHb monitoring. The latter approach has been recently further extended to include also metabolic mapping of oxCCO by Wang *et al.* [69], providing the broadest insight onto brain haemodynamics and metabolism to date. Combination of HSI mapping of intrinsic chromophores with fluorescence imaging is another presently explored multimodal fusion, as shown by preliminary work by Nardini *et al.* [43] on *ex vivo* glioma, biopsies, targeting simultaneous multiplexing of $HbO_2$, HHb, oxCCO and autofluorescence from NADH and FAD.

### 2.3. Novel methodologies to analyse the HSI data

In terms of HSI data processing and analysis, major efforts in the past years focused on improving the capability to handle the large amount of multidimensional data provided by the modality. With the sheer computational burden having increased in parallel with the augmented performances -related to spatial resolution, number of spectral bands and total length of acquisitions- that new hyperspectral technologies have been able to achieve for brain haemodynamic and metabolic monitoring (as mentioned in Sec. 2.1), novel algorithms and approaches have been explored. The driving force behind these included not only maximising data quality and achieving superior accuracy but also pushing current temporal resolution towards as close as possible to real-time HSI [70].

Artificial intelligence (AI) and machine learning (ML) methods provided enormous advantages for the aforementioned purposes, with particular relevance in the clinics [71,72]. The most predominant application to cerebral HSI has been so far towards increasing processing speed and accuracy in brain tissue features identification, specifically classification between regions of vascular vs. functional vs. cancer, as well as fine segmentation of boundaries between cerebral tumour vs. healthy tissue for intraoperative purposes [73–77]. Compared to MBLL (Sec. 2.1), most of these approaches are based on qualitative classification algorithms (either supervised or unsupervised) aimed at extracting and grouping spatio-spectral features, such as principal component analysis (PCA), optimisation algorithms, support vector machines (SVMs) and convolution neural networks (CNNs). Despite relying on differences in spectral signatures originating from known cerebral haemodynamic and metabolic processes, these methodologies still lack direct grounding to the physiological and optical states of the investigated tissue. Indeed, they tend to produce qualitative assessments (grouping brain tissue by type or function), instead of strictly quantitative ones (estimating cerebral tissue composition in terms of biomarkers), albeit with excellent accuracy and sensitivity in such classifications (even beyond 80%) [78]. Nonetheless, other computational attempts of late have also tried to bridge such gap between pure spectral components extraction and physiological/optical background knowledge with the aid of deep learning: Ezhov *et al*. [79] recently proposed a deep learning method based on MBLL for spectral unmixing of HSI targeting quantification of cerebral $HbO_2$, HHb, CCO, as well as water and lipid content, and validated it on both preclinical and clinical datasets.

Finally, as an opposite trend to enhancing data processing power of large multidimensional datasets provided by HSI, alternative approaches have instead focused on ways to reduce such dimensionality and efficiently downsize the volume of information to minimum optimal levels. Among these, it is worth mentioning the efforts spent in optimising the selection of wavelengths to use for HSI monitoring haemodynamic and metabolism in the brain, particular for targeting $HbO_2$, HHb and oxCCO simultaneously [32,33,80,81], as well as the attempts at increasing computational speed by reducing the hyperspectral problem to RGB [30,82] or dual/tri-wavelengths equivalents [83].

### 2.4. From preclinical into the hospitals: new HSI frontiers

Probably the most impactful and significant progress occurred since our 2018 review in the application of HSI to cerebral haemodynamic and metabolic monitoring was its successful advancements towards clinical translation (as already emerged clearly from discussions in the previous sections). In our original article, the majority of the works we reported were preclinical studies on animal models, with only very few applications on humans and in hospitals. Among these, almost their entirety focused on observational studies during epilepsy surgery (still explored today [84,85]). Jumping forward by merely eight years, the use of HSI in brain haemodynamic and metabolic mapping can be found solidly implemented in clinical environments, with the application to intraoperative guidance during brain tumour surgery (mainly glioma and meningioma resection) being the most explored and promising area of investigation. As a testament to this, various -and sometimes overlapping- reviews on this specific topic have been published recently [86–90], whilst three major EU-funded project have been or a currently ongoing with the goal of advancing the transition of HSI into routine clinical practice for neurooncological navigation: namely, the HELICoiD project as the forerunner [64], and recently the HyperProbe consortium [91,92] and the STRATUM project [60]. In addition to these research endeavours, commercial ventures have also started to target the field (citing again Hypervision Surgical as the most specific party involved into this from the industry).

Furthermore, other clinical areas of investigation using HSI for brain physiological imaging have emerged strongly besides tumour surgery, such as paediatric neurosurgery [93], cerebrovascular surgery [68,94,95], as well as *ex-vivo* brain cancer histopathology [43,56,57,96–98].

## 3. Conclusions and future directions

The progresses and advancements in the development and application of HSI for brain haemodynamic and metabolic monitoring since our 2018 review article have been substantial, considering the relatively short timespan interlapped. In just less than ten years since our original overview, the technique has evolved both technologically, methodologically, as well as translationally. With synergic efforts from research, academia, industry and the clinics, the application has rapidly moved from preclinical studies to the hospitals, as well as from laboratories to the market.

Novel works on HSI for preclinical, *in-vivo* and *ex-vivo*, brain haemodynamic and metabolic monitoring have strongly established it as a powerful tool for investigations in life sciences, thanks to the quantitative, fast and broad spectrum of information it can provide on cerebral oxygen physiology. This makes it particularly attractive as a compact, minimally-invasive and versatile imaging platform to study a multitude of processes involving the brain, such as neurovascular coupling, stroke, ischemia/hypoxia, tumour growth and therapeutic tracking.

Recent efforts towards clinical translation are also supporting medical HSI of brain haemodynamic and metabolic monitoring as a novel and transformative approach in healthcare, envisioning it as a cost-effective, easily-integrative, label-free, neuronavigational and diagnostic modality. It has shown the potential in many fields -particularly intraoperative brain surgery and histopathological screening- to cater for those unmet needs still present in the clinics, such as the demand for close-to-real-time streamlining of valuable data, as well as for more quantitative and accurate power. The latter are the future directions towards where the technology is currently striving, thanks to important boosts from developments in multimodal combination with other complementary techniques (e.g., LSCI and fluorescence) and from deep learning analyses.

The future of HSI for brain haemodynamic and metabolic monitoring keeps being bright, today as it was in 2018, and we believe that there is a strong rational for the technique to make a significant scientific, societal and economic impact.

## Acknowledgements

LG, FL and IT are supported by: (1) the HyperProbe project, as part of the European Union's Horizon Europe research and innovation programme under grant agreement No 101071040 and of the UKRI grant number 10048387; and (2) the fastMOT project, as part of the EU's HORIZON EUROPE programme under grant agreement number 101099291 and of the UK Research and Innovation (UKRI) under the UK government's Horizon Europe funding guarantee (grant number 10063660).

Due to the short nature of the commentary format, the authors have decided to focus on a limited number of recent publications that surely does not encompass the entire library available in the literature on the topic. The authors apologise if any specific contributions to the field have been omitted or went unnoticed in the updated review.